\documentclass[twocolumn,english,superscriptaddress,amsmath,floatfix,longbibliography,floats,prx]{revtex4-1}
\usepackage[colorlinks=true,urlcolor=blue,citecolor=blue,linkcolor=blue]{hyperref} 
\usepackage[T1]{fontenc}
\usepackage[utf8]{inputenc}
\usepackage{amssymb}
\usepackage{graphicx}
\usepackage{amsmath,color}
\usepackage{mathrsfs}
\usepackage{float}
\usepackage{indentfirst}
\usepackage{txfonts}
\usepackage[normalem]{ulem}
\usepackage[table]{xcolor}
\usepackage{array,ragged2e}
\usepackage{grffile}
\usepackage{multirow}
\usepackage{algpseudocode}
\usepackage{epstopdf}



\usepackage{babel}

\begin{document}

\hyphenpenalty=5000

\tolerance=1000

\title{Quantum Criticality with Emergent Symmetry in the Extended Shastry-Sutherland Model}

\author{Wen-Yuan Liu}
 \affiliation{Division of Chemistry and Chemical Engineering, California Institute of Technology, Pasadena, California 91125, USA}
\author{Xiao-Tian Zhang}
 \affiliation{School of Physics, Beihang University, Beijing 100191, China}
\author{Zhe Wang}
\affiliation{Department of Physics, Beijing Normal University, Beijing 100875, China}
  \author{Shou-Shu Gong}
 \affiliation{School of Physical Sciences, Great Bay University, Dongguan 523000, China, and \\
 Great Bay Institute for Advanced Study, Dongguan 523000, China}
 \author{Wei-Qiang Chen}
 \affiliation{Institute for Quantum Science and Engineering and Department of Physics, Southern University of Science and Technology, Shenzhen 518055, China}
\affiliation{Shenzhen Key Laboratory of Advanced Quantum Functional Materials and Devices, Southern University of Science and Technology, Shenzhen 518055, China}
 \author{Zheng-Cheng Gu}
 \affiliation{Department of Physics, The Chinese University of Hong Kong, Shatin, New Territories, Hong Kong, China}

\date{\today }

\begin{abstract}

Motivated by the novel phenomena observed in the layered material $\rm SrCu_2(BO_3)_2$, 
the Shastry-Sutherland model (SSM) has been extensively studied as the minimal model for $\rm SrCu_2(BO_3)_2$. However, the nature of its quantum phase transition from the plaquette valence-bond solid (PVBS) to antiferromagnetic (AFM) phase is under fierce debate, posing a challenge to understand the underlying quantum criticality. Via the state-of-the-art tensor network simulations, we study the ground state of the SSM on large-scale size up to $20 \times 20$ sites. We identify the continuous transition nature accompanied by an emergent O(4) symmetry between the PVBS and AFM phase, which strongly suggests a deconfined quantum critical point (DQCP). Furthermore, we map out the phase diagram of an extended SSM that can be continuously tuned to the SSM, which demonstrates the same DQCP phenomena along a whole critical line. Our results indicate a compelling scenario for understanding the origin of the proposed proximate DQCP in recent experiments of $\rm SrCu_2(BO_3)_2$.

\end{abstract}
\maketitle

{\it Introduction.} The frustrated Shastry-Sutherland magnet $\rm SrCu_2(BO_3)_2$  
is a fascinating material to explore exotic phenomena including magnetization plateaus, supersolid, and topological physics~\cite{plateau1999,supersolid1, SCBO2002,onizuka20001,takigawa2004, supersolid2,levy2008, sebastian2008,jaime2012,takigawa2013,matsuda2013, haravifard2016, parameswaran2013topo,romhanyi2015hall,mcclarty2017topo,wulferding2021}.
Recently, $\rm SrCu_2(BO_3)_2$ has also attracted drastically increasing interest as a promising platform to experimentally study the physics of deconfined quantum critical point (DQCP)~\cite{DQCP1,DQCP2,zayed20174,guo2020quantum}, which represents a new framework for understanding the unconventional continuous transitions beyond Landau-Ginzburg-Wilson paradigm and has been studied primarily through theoretical and numerical ways~\cite{Motrunich2004,you2016,wangchong2017,Metlitski2018,jian2018,JQ2007,JQ2008,JQ2008_2,JQ2009,loopmodel2,charrier2010,JQ2010,JQ2011,JQ2013,JQ2013_2,JQ2013_3,JQ2015,loopmodel1,JQ2016,sreejith2019,fermionDQCP2016,fermionDQCP2016,fermionDQCP2017,fermionDQCP2018,fermionDQCP2018_2,fermionDQCP2019,liuQSL,liuj1j2j3,j1xj1yj2}. 
Remarkably, the recently observed magnetic field-driven PVBS-AFM transition in $\rm SrCu_2(BO_3)_2$ displays extraordinary characteristics, including quantum critical scaling and emergence of O(3) symmetry, that strongly suggests a nearby DQCP~\cite{cui2023proximate} and calls for theoretical understanding.

\begin{figure}[tbp]
 \centering
 \includegraphics[width=3.4in]{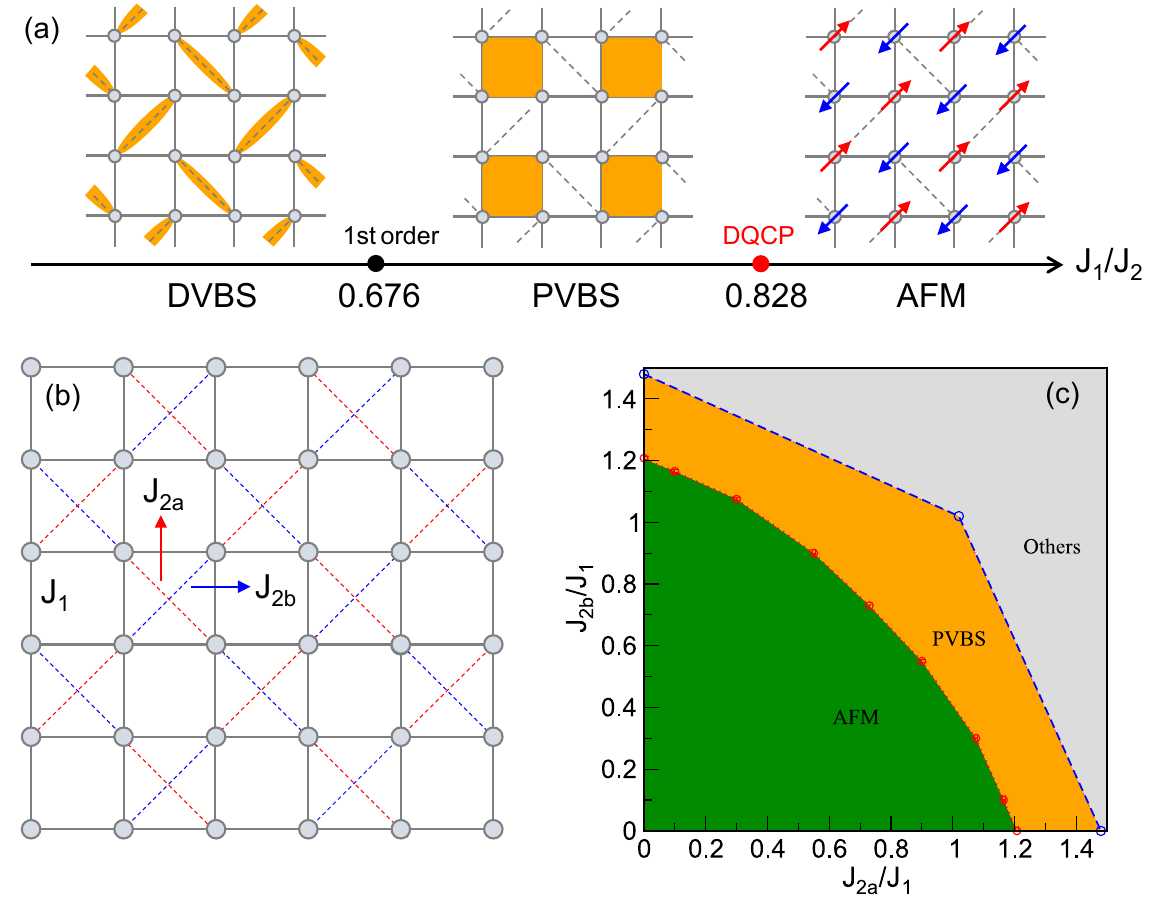}
 \caption{(a) Ground state phase diagram of the SSM, including three phase: DVBS, PVBS and AFM. Dashed diagonal lines indicate the NNN interaction terms. (b) The extended SSM, which contains two sets of NNN interaction $J_{2a}$ (red) and $J_{2b}$ (blue). (c) Ground state  phase diagram of the extended Shastry-Sutherland model and it is symmetric about $J_{2a}$ and $J_{2b}$. $J_{2b}=0$ (or $J_{2a}=0$) corresponds to the SSM, and $J_{2a}=J_{2b}$ corresponds to the checkerboard model (CBM). The grey region denotes other phases not of interest here. }
 \label{fig:phasediagram}
 \end{figure}

While the exact interactions for $\rm SrCu_2(BO_3)_2$ are still unclear, most experimental observations can be understood based on the minimal
Shastry-Sutherland model (SSM)~\cite{shastry1981exact,miyahara1999exact}, with the following Hamiltonian:
 \begin{equation}
H=J_1\sum_{\langle i,j \rangle}\mathbf{S_i}\cdot\mathbf{S_j}+J_{2}\sum_{\langle\langle
i,j\rangle\rangle}\mathbf{S_i}\cdot\mathbf{S_j},
\label{ssmodel}
\end{equation}
where $J_1$ and $J_2$ terms sum over all the nearest-neighbor (NN) [gray lines in Fig.~\ref{fig:phasediagram}(b)] and part of the next-nearest-neighbor (NNN) AFM Heisenberg interactions [red dashed lines denoted as $J_{2a}$ in Fig.~\ref{fig:phasediagram}(b)] on the square lattice.
The model is in a dimer valence-bond solid (DVBS) phase for small $J_1/J_2$ and an AFM phase for large $J_1/J_2$. 
The existence of a plaquette VBS phase (PVBS) between the DVBS and AFM phase has also been well established with a first-order DVBS-PVBS transition~\cite{corboz2013tensor,boos2019competition,lee2019signatures,Shimokawa2021signtures,xi2021first,yang2022quantum,kelecs2022rise,wang2022quantum,wang2023plaquette,zayed20174,guo2020quantum,cui2023proximate,shi2022discovery,jimenez2021quantum}. 
Nevertheless, the transition nature between the PVBS and AFM phase, which is a promising candidate of DQCP and may play a crucial role for understanding the proximate DQCP phenomena of $\rm SrCu_2(BO_3)_2$~\cite{koga2000quantum,corboz2013tensor,lee2019signatures,yang2022quantum,xi2021first,kelecs2022rise,wang2022quantum,wang2023plaquette,zayed20174,guo2020quantum,jimenez2021quantum,cui2023proximate}, remains unresolved.

Specifically, whether the potential PVBS-AFM transition belongs to first order, a DQCP, or is extended to a gapless quantum spin liquid (QSL) phase in between is a big enigma.
Series expansion study finds a continuous PVBS-AFM transition at $J_1/J_2 = 0.86(1)$~\cite{koga2000quantum,takushima2001competing}, but infinite-size tensor network  studies suggest a weakly first-order transition at $J_1/J_2 \simeq 0.78$~\cite{corboz2013tensor,xi2021first} and a multicritical point as a DQCP by introducing additional interactions~\cite{xi2021first}.
In recent density matrix renormalization group (DMRG) studies, different conclusions have also been found. 
While an earlier study suggests a direct PVBS-AFM transition, which is further argued to represent a DQCP with emergent O(4) symmetry based on quantum field theory analyses~\cite{lee2019signatures}, a recent DMRG study based on gap crossings finds an intermediate gapless QSL phase between the PVBS and AFM phase for $0.79\lesssim J_1/J_2 \lesssim 0.82$~\cite{yang2022quantum}.
This PVBS-QSL-AFM picture is also supported by a functional renormalization group study~\cite{kelecs2022rise}. 
These controversial results lead to a big obstacle for further understanding of the proximate DQCP phenomena in $\rm SrCu_2(BO_3)_2$ observed in a magnetic field and high pressure~\cite{cui2023proximate}.
 
In this paper, we employ the state-of-the-art tensor network simulations by means of the finite projected entangled pair state (PEPS) approach to study an extended SSM with the following Hamiltonian:
\begin{equation}
H=J_1\sum_{\langle i,j \rangle}\mathbf{S_i}\cdot\mathbf{S_j}+J_{2a}\sum_{\langle\langle
i,j\rangle\rangle}\mathbf{S_i}\cdot\mathbf{S_j}+J_{2b}\sum_{\langle\langle
i,j\rangle\rangle}\mathbf{S_i}\cdot\mathbf{S_j},
\label{ss_cbmodel}
\end{equation}
where $J_{2a}$ and $J_{2b}$ represent two sets of the NNN interactions as shown in Fig.~\ref{fig:phasediagram}(b). 
For either $J_{2b}=0$ or $J_{2a}=0$, the model recovers the standard SSM, while for $J_{2a}=J_{2b}$ it is the checkerboard model (CBM) with a potential PVBS-AFM transition~\cite{starykh2005anisotropic, bishop2012frustrated,zou2020nearly}.
The two models can be connected by tuning $J_{2a}$ and $J_{2b}$.
Our extensive tensor network simulations yield strong evidence that the SSM undergoes a direct PVBS-AFM transition accompanied by an emergent O(4) symmetry, supporting a DQCP in the SSM. The conclusions are further strengthened by the investigations of the extended SSM, which exhibits the same type of DQCP physics as shown in the phase diagram Fig.~\ref{fig:phasediagram}(c). Our findings suggest an origin of the experimentally observed proximate DQCP phenomena in $\rm SrCu_2(BO_3)_2$. 

{\it Results of the SSM.} The finite PEPS algorithm here we adopt has been demonstrated as a powerful approach for simulating quantum spin systems~\cite{liu2017,liu2018,liufinitePEPS,liuQSL,liuj1j2j3,j1xj1yj2}. To further verify the finite PEPS results, we carefully compare  with  DMRG and infinite-size tensor network methods on the SSM~\cite{corboz2013tensor,xi2021first}. These results explicitly demonstrate the high accuracy of the finite PEPS method. ( In addition, we also benchmark our results of $J-Q$ model with Quantum Monte Carlo simulation at all different sizes. See Supplemental Material~\cite{SM_ssmodel}.) We use $D=8$ to perform all calculations, where both the ground state energy and order parameters are well converged for the largest available size  $20\times 20$ by very careful analysis for $D=4-10$ results~\cite{SM_ssmodel}.

 \begin{figure}[tbp]
 \centering
 \includegraphics[width=3.4in]{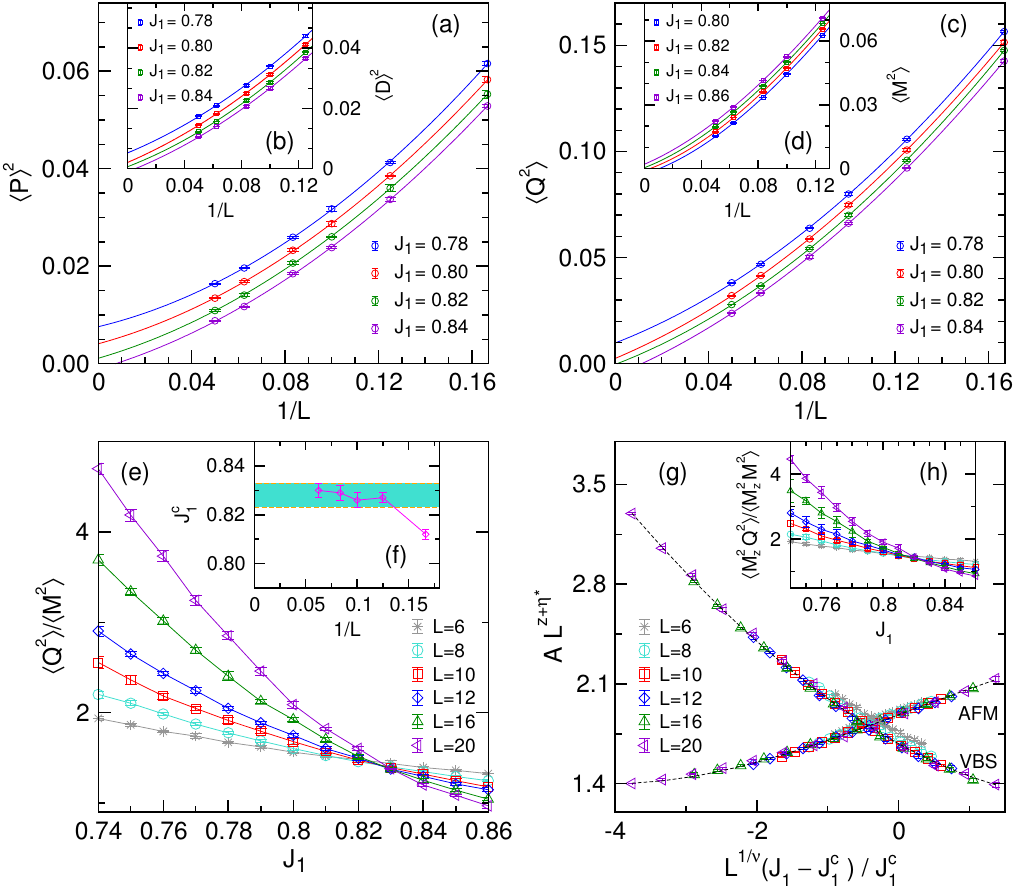}
 \caption{  Finite size scaling of ground state order parameters for the SSM on $L\times L$ lattices (a)-(d). Quadratic fits are shown for VBS order parameters (a-c) and cubic fits are shown for AFM order parameter (d) using $L=6-20$. The  ratios of PVBS and AFM order parameters $\langle Q^2\rangle/\langle {\bf M}^2\rangle$ are presented in (e) and their crossing points from $(L ,L+2)$ or $(L, L+4)$ are presented in (f). (h) shows the fourth-order momentum ratio. (g) shows the data collapse of PVBS and AFM order parameters, and the dashed lines are  quadratic curves using the corresponding critical exponents. The values of scaled AMF quantities in (g) have been shifted upwards by 0.7 for clear display.}
 \label{fig:SSmodel}
 \end{figure}

Since our method works well for open boundary systems, we first look into the boundary induced dimerized pattern on $L\times L$ systems for the SSM. 
We set $J_2=1$ to keep the same energy scale with previous studies.  The local plaquette operator at site ${\bf i}=(i_x,i_y)$ is defined as $\Pi_{{\bf i}}=\frac{1}{2}(P_{\square,{\bf i}}+P^{-1}_{\square,{\bf i}})$, where $P_{\square,{\bf i}}$ denotes the cyclic exchange operator of the four spins on a given plaquette at site ${\bf i}$.  The boundary induced plaquette order parameter is measured by $P=\frac{1}{N_p}\sum_{{\bf i}}(-1)^{i_x}\Pi_{{\bf i}}$, where $N_p=(L-1)^2$ is the total number of counted plaquettes~\cite{zhao2019symmetry}. The scaling of $\langle P\rangle^2$ with $L=6-20$ is shown in Fig.~\ref{fig:SSmodel}(a) for different couplings $J_1$ using second order polynomial fits. With $J_1$ increasing, $\langle P \rangle^2$ gradually decreases. The extrapolated  $\langle P\rangle^2$ in the thermodynamic limit at $J_1=0.78$, 0.80, 0.82 are 0.0075(3), 0.0041(3) and 0.0014(5), respectively, but would be zero at $J_1 =0.84$. This suggests the PVBS phase vanishes around $J_1=0.83$. The third order fits give the same conclusion. The plaquette pattern in the PVBS phase is shown in the middle part of Fig.~\ref{fig:phasediagram}(a), where the singlets form on the empty plaquettes.

On the other hand, we consider the dimer order parameters for double check, defined as 
\begin{equation}
D_{\alpha}=\frac{1}{N_b}\sum_{{\bf i}}(-1)^{i_{\alpha}}{\bf S}_{{\bf i}} \cdot {\bf S}_{{\bf i}+{\rm e_\alpha}},
\end{equation} 
where $\alpha=x$ or $y$, and $N_b=L(L-1)$ is the corresponding total number of counted bonds along the $\alpha$ direction. For the PVBS phase, the boundary induced dimerization $\langle D \rangle ^2=\langle D_x \rangle ^2+\langle D_y \rangle ^2$ should also be finite in the thermodynamic limit. The scaling of $\langle D\rangle^2$ for $J_1=0.78$, 0.80, 0.82, 0.84 using $L=6-20$ is shown in Fig.~\ref{fig:SSmodel}(b), and indeed is consistent with a PVBS phase for $J_1\lesssim 0.83$.

Now we turn to the PVBS and AFM order parameters defined based on correlation functions. The PVBS order parameter is defined as $\langle Q^2 \rangle=\langle (D_x-D_y)^2 \rangle$ where $Q=D_x-D_y$~\cite{lee2019signatures}. In Fig.~\ref{fig:SSmodel}(c), the quantity  $\langle Q^2 \rangle$ is presented. The finite-size scaling shows $J_1=0.8$ has a nonzero PVBS order 0.0024(6) in the thermodynamic limit,  and the vanishing point $J_1$ of PVBS order parameter is very close to  that obtained from the scaling of boundary induced quantities $\langle P \rangle^2$ and $\langle D \rangle^2$. On the other hand, the AFM order parameter $\langle {\bf M}^2 \rangle$ where ${\bf M}=\frac{1}{L^2}\sum_{\bf i}(-1)^{i_x+i_y}{\bf S_i}$ is show in Fig.~\ref{fig:SSmodel}(d). One can see with $J_1$ increasing,  the AFM order increases, and potentially establishes for $J_1\geq0.84$.  These results from different physical quantities suggest a direct PVBS-AFM phase transition. By using the crossing points of order parameter ratio, which we will mention later, it gives a PVBS-AFM transition point at $J_{1}\simeq 0.83$, consistent with the finite size scaling analysis.  The physical quantities vary smoothly about $J_1$ in our results, indicating a continuous PVBS-AFM transition, though the possibility of weakly first order transition in the thermodynamic limit can not be fully excluded. 

The  PVBS-AFM transition point in the  SSM has been argued as a DQCP with emergent O(4) symmetry  through the rotation of  three components of AFM order parameter ${\bf M}=(M_x, M_y, M_z)$ and one-component PVBS order parameter $Q$ into each other to form a superspin ${\bf n}=(M_x, M_y, M_z, Q)$, where  $Q=D_x-D_y$~\cite{lee2019signatures}. To check the O(4) symmetry, as we note the SO(3) symmetry for AFM components is well satisfied by observing $\langle  M^2_\alpha \rangle=\frac{1}{3} \langle {\bf M}^2\rangle$ (see SM~\cite{SM_ssmodel}), we only need to check the additional symmetry formed by $M_z$ and $Q$. A simple but nontrivial quantity to verify the emergent O(4) symmetry is that the ratio $\langle Q^2\rangle/\langle {\bf M}^2\rangle$ (or $\langle Q^2\rangle/\langle  M_z^2\rangle$  as $\langle  M^2_z \rangle=\frac{1}{3} \langle {\bf M}^2\rangle$) should be size-independent at the transition point~\cite{loopmodel2,sreejith2019,nahum2019}. In Fig.~\ref{fig:SSmodel}(e), the quantities $\langle Q^2\rangle/\langle {\bf M}^2\rangle$ for different system size have a crossing point at $J_1\simeq 0.83$, which is intrinsically close to the PVBS-AFM transition point  evaluated by finite size scaling of order parameters. The crossing points show much smaller finite size effects [Fig.~\ref{fig:SSmodel}(f)], which has also been observed in other DQCP studies by Quantum Monte Carlo simulations~\cite{loopmodel2,sreejith2019,nahum2019}. These results on one hand  provide direct evidence to support the PVBS-AFM transition point as a DQCP with emergent O(4) symmetry, on the other hand give a more accurate evaluation of the transition point $J^c_1=0.828(5)$,  significantly reducing the  uncertainties from the finite size extrapolations of order parameters. We also consider higher order momentum ratios of order parameters to check the emergent symmetry~\cite{loopmodel1,sreejith2019,nahum2019}, which is very challenging to compute. Within our capability, we can only get the  fourth-order momentum ratio $\langle M_z^2 Q^2\rangle/\langle M_z^2 {\bf M}^2\rangle$. Remarkably, $\langle M_z^2 Q^2\rangle/\langle M_z^2 {\bf M}^2\rangle$ also has a crossing point around  $J_1\simeq 0.83$, shown in Fig.~\ref{fig:SSmodel}(h), further supporting the emergent symmetry.

The extensive data with different system size enable us to precisely extract the corresponding critical exponents. We use the standard finite size scaling formula without subleading corrections~\cite{FSS1}: 
\begin{equation}
A(J_1,L)=L^{-(z+\eta^{*})}F[L^{1/\nu}(J_1-J^c_{1})/J^c_{1}], 
\label{eq:scaling}
\end{equation}
where $J_1$ denotes the coupling constant, and $A(J_1,L)$ denote the AFM order parameters  $\langle {\bf M}^2 \rangle$ or VBS order parameters $\langle Q^2\rangle$ for different size $L$ at different $J_1$.  $\nu$ is the correlation length exponent, and $z+\eta^{*}_{s,p}$ represents the spin or plaquette exponents where $z$ is the dynamical exponent and $\eta^{*}_{s,p}$ are corresponding anomalous scaling dimensions. $F[x]$ is a polynomial function. The analysis of data collapse can be followed in Ref.~\cite{liuj1j2j3}.
In  Fig.~\ref{fig:SSmodel}(g), we show the data collapse of the order parameters (using $J^c_{1}=0.828$) by collectively fitting $A(J_1,L)$ with different $J_1$ and $L$ (exclude $L=6$), and the physical quantities can be well scaled with  $z+\eta^{*}_{s}=1.39(2)$, $z+\eta^{*}_{p}=1.39(1)$ and $\nu=0.84(3)$. On the other hand, note at the critical point $J_1=J^c_1$ it has $A(J^c_1,L)\propto L^{-(z+\eta^{*})}$, indicating that the exponents $(z+\eta^{*})$ can be directly evaluated by using the data at $J^c_1$. Indeed at $J_1=0.83$ it gives the almost same value 1.39 for both $z+\eta^{*}_{s}$ and $z+\eta^{*}_{p}$, in good agreement with the collective fittings. Note that the obtained critical exponents clearly indicate $z+\eta^{*}_{s}=z+\eta^{*}_{p}$, which is consistent with the emergent O(4) symmetry~\cite{loopmodel1,nahum2019,j1xj1yj2}.

 \begin{figure}[tbp]
 \centering
 \includegraphics[width=3.4in]{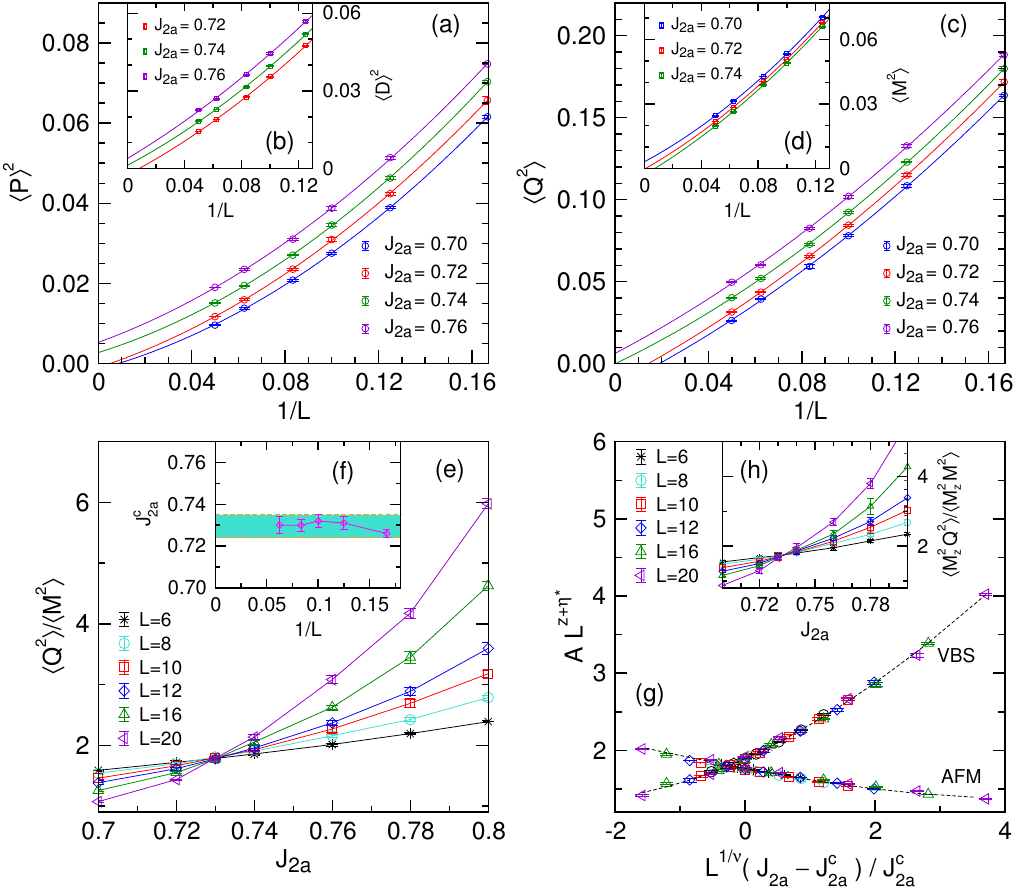}
 \caption{  Finite size scaling of ground state order parameters for  checkerboard  model ($J_{2a}=J_{2b}$) on $L\times L$ lattices (a)-(d), setting $J_{1}=1$. Quadratic fits for $L=6-20$ are shown. The meanings of the figures are the same with those of Fig.~\ref{fig:SSmodel}. The values of scaled AMF quantities in (g) have been shifted upwards by 0.7 for clear display.}
 \label{fig:CBmodel}
 \end{figure}

At last, we compare our results with those from DMRG on smaller system size.
Although converged DMRG results are unbiased, the limited system size and different analyzed quantities may lead to different conclusions for 2D limit.
By computing the ground-state energy and correlation lengths on infinite-long cylinders, an earlier DMRG calculation finds a PVBS-AFM transition that can be consistent with an emergent O(4) symmetry.
Nonetheless, a latter DMRG study using gap crossings
suggests an intermediate QSL for $g_{c1} \lesssim J_1/J_2 \lesssim g_{c2}$ separating the PVBS and AFM phase in the SSM, where $g_{c1}\simeq  0.79$ and $g_{c2}\simeq 0.82$ are estimated from the $1/L^2$ finite-size scaling of gap crossings~\cite{yang2022quantum}. Notably, the extrapolated results based on the relatively small cylinder width near transition point may strongly depend on the assumed scaling formula~\cite{qian2023}. Indeed, the extrapolated $g_{c1}$ from linear fit of $1/L$, can become very close to the extrapolated $g_{c2}$ (see SM~\cite{SM_ssmodel}). Hence it is possible that $g_{c1}$ and $g_{c2}$ would eventually converge to the same value at larger size.
Our extensive results, utilizing larger system sizes, reveal the emergence of O(4) symmetry at the point $J^c_{1}=0.828(5)$ in the SSM and similar features in the extended SSM, providing strong evidence for a direct PVBS-AFM transition and significantly diminishing the possibility of an intermediate QSL phase.

{\it Results of the extended SSM.} To gain more insight into the ground state phase diagram of SSM,  we further consider the extended SSM Eq.~\eqref{ss_cbmodel}, which is described by a two dimensional space of parameters $(J_{2a}/J_1, J_{2b}/J_1)$. 
For $J_{2b}=0$ it recovers the SSM (relabel $J_{2a}$ as $J_2$), and for $J_{2b}=J_{2a}$ it corresponds to the  CBM. The two cases can be  continuously connected by tuning the strength of $J_{2b}$ terms.  For the CBM, previous studies suggest it could have a direct PVBS-AFM transition~\cite{starykh2005anisotropic,bishop2012frustrated,zou2020nearly}. Here we focus on the region  close to the suggested PVBS-AFM transition. In Fig.~\ref{fig:CBmodel}, we present  different quantities including boundary induced orders $\langle P\rangle^2 $ and $\langle D \rangle^2$, and AFM and PVBS order parameters $\langle {\bf M}^2 \rangle$  and $\langle Q^2 \rangle$, as well as the order parameter ratios, with the systems up to $20\times 20$ sites. Similar to the analyses of the SSM, we find a continuous PVBS-AFM transition at $J_{2a}/J_1=J_{2b}/J_1\simeq 0.73$ for the CBM, associated with the emergent O(4) symmetry.

 \begin{figure}[tbp]
 \centering
 \includegraphics[width=3.4in]{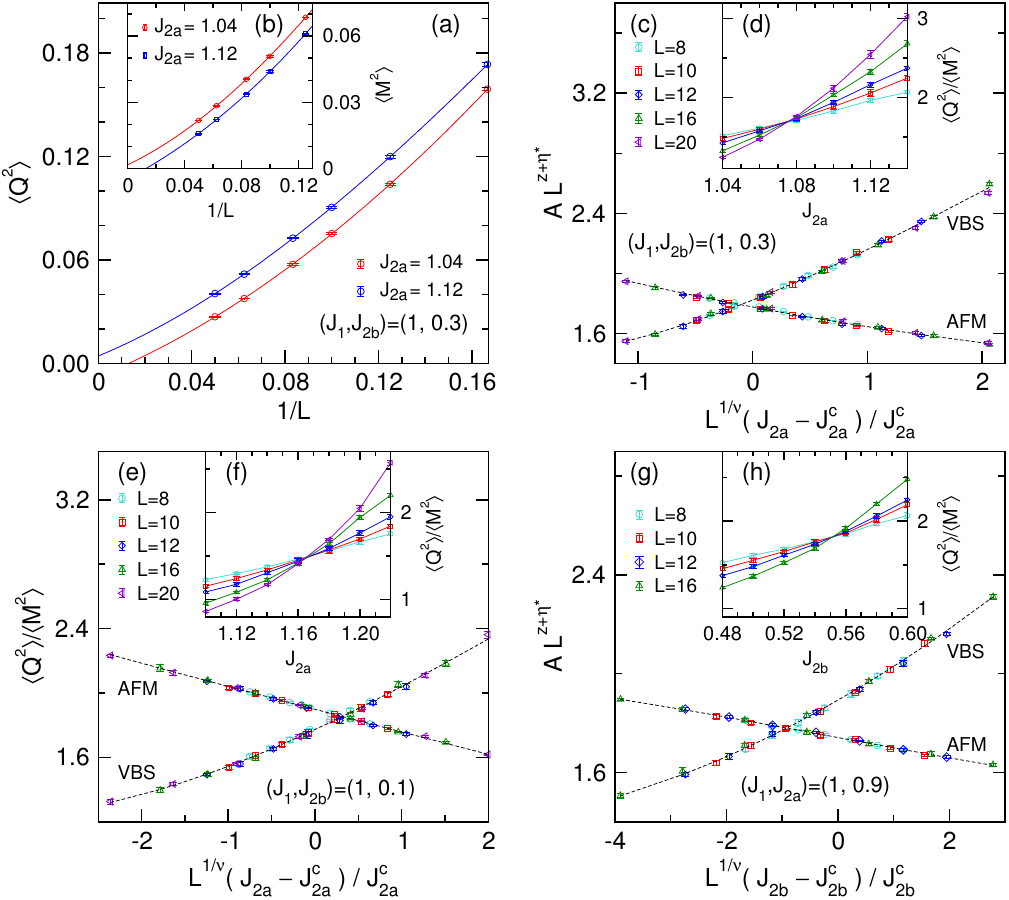}
 \caption{  The scaling of AFM and VBS order parameters for three cases: $(J_1,J_{2b})=(1,0.3)$ with tuning parameter $J_{2a}$ (a)-(d), and extrapolations with cubic fits for $L=6-20$ are shown (a-b);  $(J_1,J_{2b})=(1,0.1)$ with tuning parameter $J_{2a}$ (e)-(f); and $(J_1,J_{2a})=(1,0.9)$ with tuning parameter $J_{2b}$ (g)-(h). The values of scaled AMF quantities in (c), (e) and (g) have been shifted upwards by 0.7 for clear display. }
 \label{fig:MIXmodel}
 \end{figure}

In fact, by computing  the intermediate cases with proper $J_{2a}/J_1$ and $J_{2b}/J_1$  in between the SSM and the CBM, we find these different cases also support a continuous PVBS-AFM transition with emergent O(4) symmetry.  In Fig.~\ref{fig:MIXmodel}, we present the scaling for the cases with fixed $J_{2a}/J_1=0.9$ ( sweeping $J_{2b}$), fixed  $J_{2b}/J_1=0.1$ (sweeping $J_{2a}$),  as well as fixed  $J_{2b}/J_1=0.3$ (sweeping $J_{2a}$). In all of these cases, the PVBS has the same plaquette pattern with the SSM that the plaquette singlets are located on four sites without diagonal interaction terms [see Fig.\ref{fig:phasediagram}(b)]. The extracted critical exponents for these different cases are  given in Table.~\ref{tab:criticalexponents}, and  it clearly shows that the PVBS-AFM transitions have the similar critical exponents, $z+\eta^{*}_s=z+\eta^{*}_p \sim 1.35$ and $\nu\sim 0.80$, suggesting all of them  belong to the same universality class. The same value of $z+\eta^{*}_s$ and $z+\eta^{*}_p$ is a reflection of the emergent O(4) symmetry. Since the SSM can be smoothly connected to the CBM via the above intermediate cases,
the results of extended SSM consolidate the existence of a DQCP.

   \begin{table}[htbp]
   \centering
 \caption { Critical exponents of the  extended SSM  at the PVBS--AFM transition points. Critical points are from crossing points of the ratio of order parameters. Errors in critical exponents are from fittings. }
	\begin{tabular*}{\hsize}{@{}@{\extracolsep{\fill}}lcccc@{}}
		\hline\hline
	   model  &   $z+\eta_s^*$ & $z+\eta_p^*$    & $\nu$ & $J_c$  \\ \hline
 		
        SSM [$J_{2a}=1,J_{2b}=0$] & 1.39(2) & 1.39(1) & 0.84(3)  & $J_1=0.828(5)$\\
        CBM [$J_1=1$, $J_{2a}=J_{2b}$] & 1.33(1) & 1.33(1) & 0.82(4) &$J_{2a}= 0.730(4)$ \\
                $(J_1,J_{2b})=(1,0.1)$  & 1.39(1) & 1.39(1) & 0.80(4) & $J_{2a}=1.165(3)$\\  
             $(J_1,J_{2b})=(1,0.3)$  & 1.35(1) & 1.35(1) & 0.85(3) & $J_{2a}=1.075(4)$\\  
         $(J_{1},J_{2a})=(1,0.9)$  & 1.33(1) & 1.33(1) & 0.81(5) &  $J_{2b}=0.550(3)$\\  \hline
 		\hline\hline
	\end{tabular*}
\label{tab:criticalexponents}	
\end{table}

{\it Summary and discussion.} The nature of the PVBS-AFM transition in the SSM is a critical issue to be addressed. By applying the state-of-the-art finite-size tensor network simulations, we reveal a direct continuous PVBS-AFM transition in the SSM, accompanied by an emergent O(4) symmetry. Further investigations on the extended SSM 
find the similar behaviors. These results are consistent with the DQCP nature of PVBS-AFM transition in the SSM, and exclude the existence of an intermediate QSL phase.

In the SSM, our PVBS-AFM transition point $J^c_{1}=0.828(5)$ is a bit different from the one $J^c_1\simeq 0.78$ obtained by the infinite-size tensor network simulations using a finite bond dimension $D$~\cite{corboz2013tensor,xi2021first}. This difference in principle could be further reduced by properly extrapolating the infinite-size results with a new finite correlation length scaling scenario or using a sufficiently large $D$
~\cite{liuj1j2j3,corboz2018,rader2018,Vanhecke2022scaling,liao2017}.
While the PVBS-AFM transition in our large-scale simulation is extremely close to a continuous transition, the possibility of a weakly first-order transition in the thermodynamic limit still cannot be fully excluded.

In the PVBS phase of $\rm Sr{Cu}_2{(BO_3)}_2$, the plaquette singlets form at the sites with diagonal interactions, which is different from the SSM and is possibly owing to the weak intralayer and interlayer interactions~\cite{DM1999,DM2004,zayed20174,guo2020quantum,cui2023proximate,boos2019competition,jimenez2021quantum,xi2021first}.
Interestingly, in the phase diagram of $\rm Sr{Cu}_2{(BO_3)}_2$, including these extra interactions, the PVBS-AFM transition line in $\rm Sr{Cu}_2{(BO_3)}_2$ still may connect the nearby DQCP in the SSM via a triple critical point~\cite{xi2021first}, exhibiting large correlation lengths with emergent symmetry~\cite{lee2019signatures}. 
On the other hand, the PVBS in $\rm Sr{Cu}_2{(BO_3)}_2$ also has two-fold degeneracy, which in presence of magnetic fields also naturally supports the observed O(3) symmetry. Thus, our findings strongly suggest that it is a highly compelling scenario that the proximate DQCP phenomena observed in $\rm Sr{Cu}_2{(BO_3)}_2$ under magnetic fields, including the emergent O(3) symmetry and quantum critical scaling at the PVBS-AFM transition~\cite{cui2023proximate}, can originate from the DQCP universality class  inherent in the SSM~\cite{lee2019signatures}. Consequently,  the anomalous dimension scaling $\eta \sim 0.2$ under magnetic fields in $\rm Sr{Cu}_2{(BO_3)}_2$~\cite{cui2023proximate} is understandably near to our result $\eta\sim 0.35$ without magnetic fields (setting the dynamic exponent $z=1$). Note that the SSM subjected to magnetic fields and additional interactions in $\rm Sr{Cu}_2{(BO_3)}_2$, e.g., frustrated bilayer interaction $\sim 0.1J_2$, Dzyaloshinskii–Moriya interaction $\sim 0.03J_2$, and higher-order further-neighbour interaction~\cite{DM1999,DM2004,jimenez2021quantum}, can posses very rich  physics~\cite{DM2001,SS2003,Kodama2005, supersolid2,DM2011,SS2014, romhanyi2015hall,mcclarty2017topo,shi2019emergent,shi2022discovery,nomura2023unveiling,wang2023plaquette,jimenez2021quantum,cui2023proximate}, and we leave these interesting topics to future studies. 
\\

{\it Acknowledgments.} We thank Philippe Corboz, Wenan Guo, Wei Li, Fr\'ed\'eric Mila, Junsen Wang, Ning Xi, Zhi-Yuan Xie and Rong Yu for very helpful discussions. We also thank  Philippe Corboz and Ning Xi for providing the iPEPS and iPESS data, respectively.  This work was supported by the CRF C7012-21GF from the Hong Kong’s Research Grants Council, and the National Key R\&D Program of China (Grants No. 2022YFA1403700). W.Q.C. was supported by the National Natural Science Foundation of China (Grants No. 12141402), the Science, Technology and Innovation Commission of Shenzhen Municipality (No. ZDSYS20190902092905285), and the Guangdong Basic and Applied Basic Research Foundation under Grant No. 2020B1515120100. This work was also granted access to the HPC resources of  Center for Computational Science and Engineering at Southern University of Science and Technology. W.Y.L. was  supported by the U.S. Department of Energy, Office of Science, National Quantum Information Science Research Centers, Quantum Systems Accelerator. X.T.Z. and S.S.G. were supported by the National Natural Science Foundation of China (No. 11874078 and No. 11834014), the Special Project in Key Areas for Universities in Guangdong Province (No. 2023ZDZX3054), and the Dongguan Key Laboratory of Artificial Intelligence Design for Advanced Materials. Z.W. was supported by the National Natural Science Foundation of China under Grant No. 12175015.

\clearpage

\appendix
\setcounter{equation}{0}
\newpage

\renewcommand{\thesection}{S-\arabic{section}} \renewcommand{\theequation}{S%
\arabic{equation}} \setcounter{equation}{0} \renewcommand{\thefigure}{S%
\arabic{figure}} \setcounter{figure}{0}

\centerline{\textbf{Supplemental Material}}

\section{Tensor Network Method}

  \begin{table*}[htbp]
\caption { Comparison between PEPS and QMC results on $J-Q$ model on open boundary systems, including ground state energies (set $Q=1$),  magnetizations $\langle M^2_z\rangle$ and dimerizations $\langle \Bar{D} \rangle^2$. In PEPS calculations, Monte Carlo sampling errors are order of $10^{-6}$ for energies $E_{PEPS}$ and order of $10^{-5}$ for $\langle M^2_z\rangle$ and $\langle \Bar{D} \rangle^2$. For QMC results $\beta=120$ is used, and sampling errors for these quantities are order of  $10^{-6}$. }

	\begin{tabular*}{\hsize}{@{}@{\extracolsep{\fill}}lllllll|llllll@{}}
\hline\hline
   \multicolumn{1}{c}{\multirow{1}{*}{}}
   &\multicolumn{1}{c} {\multirow{1}{*}{$J/Q=0$}}
   &\multicolumn{1}{c} {\multirow{1}{*}{}}
      &\multicolumn{1}{c} {\multirow{1}{*}{}}
    &\multicolumn{1}{c} {\multirow{1}{*}{}}
         &\multicolumn{1}{c} {\multirow{1}{*}{}}
    &\multicolumn{1}{c|} {\multirow{1}{*}{}}
        &\multicolumn{1}{c} {\multirow{1}{*}{$J/Q=0.1$}} 
    &\multicolumn{1}{c} {\multirow{1}{*}{}}
          &\multicolumn{1}{c} {\multirow{1}{*}{}}
    &\multicolumn{1}{c} {\multirow{1}{*}{}}
             &\multicolumn{1}{c} {\multirow{1}{*}{}}
    &\multicolumn{1}{c} {\multirow{1}{*}{}} \\
       \hline   
    $L$ & $E_{PEPS}$ & $E_{QMC}$ & $\langle M^2_z\rangle_{PEPS}$ & $\langle M^2_z\rangle_{QMC}$ & $\langle \Bar{D}\rangle^2_{PEPS}$& $ \langle \Bar{D} \rangle^2_{QMC}$ & $E_{PEPS}$ & $E_{QMC}$ & $\langle M_z^2\rangle_{PEPS}$ & $\langle M_z^2\rangle_{QMC}$ & $\langle \Bar{D}\rangle^2_{PEPS}$& $ \langle \Bar{D}\rangle^2_{QMC}$
      \\
   \hline

 	    6 & -0.66172 & -0.66174 &0.02531(8) &0.02539& 0.00572(2)&0.00565   & -0.71708 & -0.71711 & 0.02665(6) &0.02672& 0.00533(2)&0.00526\\
		8 & -0.70379 & -0.70383 &0.01565(9) &0.01573& 0.00433(2)&0.00430    & -0.76015 & -0.76019 & 0.01670(3)&0.01684& 0.00396(1)&0.00392\\
		10 & -0.72897 & -0.72903 &0.01074(6)&0.01092& 0.00355(1)&0.00349   & -0.78603 & -0.78608 & 0.01175(3)&0.01187& 0.00316(0)&0.00312\\
 		12 & -0.74558 & -0.74564 &0.00804(4)&0.00812& 0.00298(3)&0.00295   & -0.80318 & -0.80322 & 0.00890(4)&0.00898& 0.00262(1)&0.00258\\
 		14 & -0.75735 & -0.75739  &0.00633(2)&0.00634& 0.00260(1)&0.00256  & -0.81534 & -0.81539 & 0.00704(4) &0.00712& 0.00224(2)&0.00219\\
       	16 & -0.76606 & -0.76613  &0.00507(3)&0.00513& 0.00231(2)&0.00227  & -0.82440 & -0.82446 & 0.00577(3)&0.00585& 0.00194(1)&0.00191 \\
            20 & -0.77812 & -0.77820  &0.00353(2)&0.00360& 0.00192(1)&0.00186  & -0.83694 & -0.83703& 0.00412(3)& 0.00424& 0.00157(1)&0.00151\\
             \hline\hline
	\end{tabular*}
	
\label{tab:jqmodel}	
\end{table*}

The tensor network state, particularly the projected entangled pair state (PEPS), serves as a highly potent description for quantum many-body systems~\cite{verstraete2008}. It has been extensively employed to characterize complex entangled many-body states, including those with topological order~\cite{cirac2021matrix}. The expressive capacity of PEPS is systematically governed by the dimension $D$ of tensors. In this study, we adopt the tensor network formalism using finite PEPS in the framework of variational Monte Carlo~\cite{liu2017,liufinitePEPS}, which works very well to deal with open boundary systems. This approach has proven very successful in elucidating the nature of particularly challenging frustrated spin systems~\cite{liuQSL,liuj1j2j3,j1xj1yj2}. The PEPS ansatz here we used is always defined on a square lattice with open boundaries. The tensors on the corners, edges and in the middle region, have 2, 3 and 4 virtual indices, respectively, to connect their nearest neighbour sites. Throughout this paper, unless otherwise specified, we use PEPS with $D=8$ for all calculations which can provide well-converged results (see Sec.~\ref{app:convergence}). Here we show more results to further demonstrate the accuracy of our method.

\subsection{comparison with QMC on $J-Q$ model}

Here we present a comprehensive benchmark study on the spin-1/2 $J-Q$ model, also known as the Heisenberg model with additional four-spin exchange interactions~\cite{JQ2007}. This model can be unbiasedly simulated using quantum Monte Carlo (QMC) without encountering sign problems and served as a representative example for studying the DQCP~\cite{JQ2007}. The $J-Q$ model has a VBS-AFM phase transition at $J/Q=0.0447(2)$~\cite{sandvik2012finite}. 
Due to the existence of four-spin interactions and critical properties for the $J-Q$ model, it thus also offers a highly nontrivial example to demonstrate the capability of PEPS to deal with critical systems.

Here we focus on two commonly studied  near critical points: $J/Q=0$ and $J/Q=0.1$, which have very large correlation lengths around $\xi\sim 30$~\cite{sandvik2012finite}. The QMC results are obtained in the stochastic series expansion (SSE) method~\cite{FSS1} on open boundary conditions, which operates in the $S_z$ basis. In this context it is convenient to calculate the $z$-component quantities, including staggered magnetization $\langle M^2_z\rangle=\frac{1}{N^2}\langle (\sum_{\bf i}(-1)^{i_x+i_y} S^z_{\bf i})^2 \rangle$ where $N=L^2$, as well as the boundary-induced dimerization $\langle \Bar{D} \rangle^2=\frac{1}{2}(\langle \Bar{D}_x\rangle^2+\langle \Bar{D}_y \rangle^2)$, where $\Bar{D}_{\alpha}=\frac{1}{N_b}\sum_{{\bf i}}(-1)^{i_{\alpha}}{S}^z_{{\bf i}} {S}^z_{{\bf i}+{\rm e_\alpha}}$ with $N_b=L(L-1)$.  

For the PEPS simulation, we use $D=8$ PEPS as the variational ansatz, and stochastic gradient descent~\cite{liu2017,liufinitePEPS} and stochastic reconfiguration~\cite{Vieijra2021direct} to  optimize the PEPS wave function. After optimization, we can compute the  physical quantities directly via Monte Carlo sampling of the optimized PEPS. Such energies can also be compared with the ones computed from the wave function by symmetrizing the optimized PEPS with horizontal and vertical lattice reflection and spin inversion symmetry~\cite{liufinitePEPS}, which are listed in Table~\ref{tab:jqmodel}. For $L=20$, the energies using symmetrization  at $J/Q=0$ and 0.1 are $-0.77812$ and $-0.83694$, respectively, slightly lower than -0.77806 and -0.83688 without symmetrization, with a difference around 0.00006. For $L\leq 16$, with and without symmetrization, the energy differences are also as small as 0.00002. In all these cases, we find magnetization and dimerization have no visible changes with and without symmetrization, which is similar to the study of the Heisenberg model~\cite{liufinitePEPS}. The comparison about symmetrization indicates our obtained PEPS is well optimized as a good ground state that can well respect the lattice and spin symmetry. 

In Table~\ref{tab:jqmodel}, we summarize  the physical quantities obtained from QMC and PEPS.  It is clear that the obtained physical quantities including energies, magnetization, and dimerization from PEPS are in excellent agreement with those from QMC across different system sizes, and the energy differences can be as small as 0.00009 on $20\times 20$ sites. These results show  that $D=8$ PEPS can produce well converged results  up to $20\times 20$ sites, and  demonstrate the extraordinary capability of PEPS in handling near critical systems.

\subsection{comparison with DMRG and convergence}
\label{app:convergence}

 \begin{figure}[htbp]
 \centering
 \includegraphics[width=3.4in]{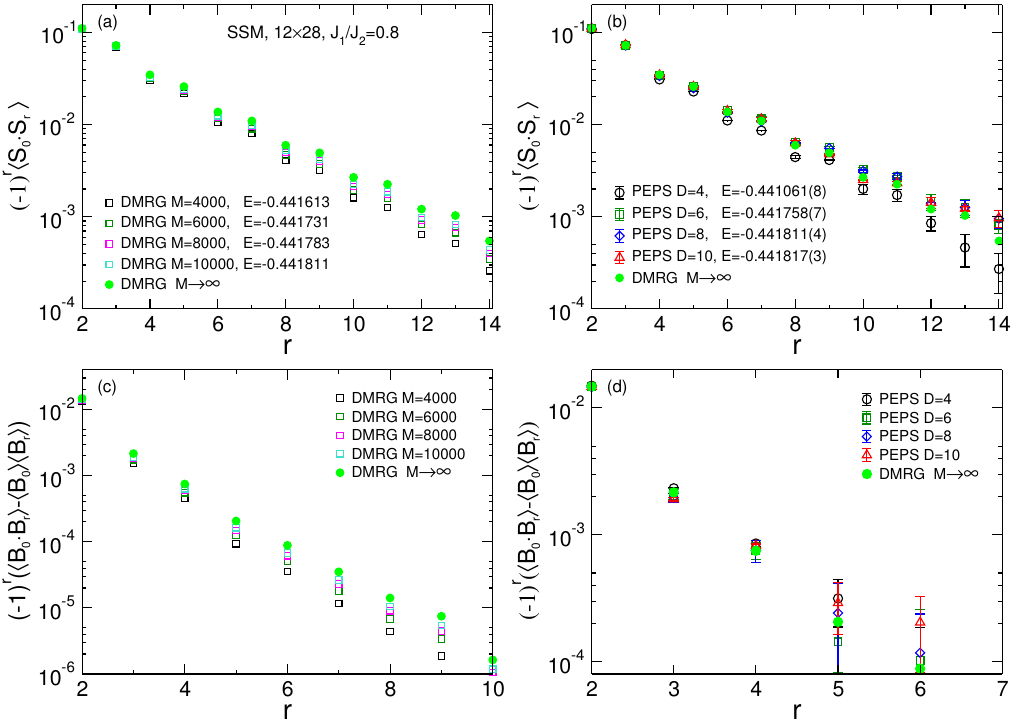}
 \caption{Ground-state energy, spin, and dimer correlation functions from the DMRG and PEPS simulations on the $12 \times 28$ SSM at $J_1/J_2=0.8$. $M$ is the number of SU(2) multiplets in the DMRG simulations with SU(2) symmetry. The correlation functions are measured along the central row $y=6$. The reference site for spin correlation is chosen at the third column near the left edge. The dimer correlations are measured for the nearest-neighbor bonds along the $x$ direction with the reference bond located between the third and fourth columns.}
 \label{fig:DMRG}
 \end{figure}

Now we directly compare the PEPS and DMRG results of the SSM on a long $L_y\times L_x$ strip with $L_y=12$, $L_x=28$. Here we focus on $J_1/J_2=0.8$ (set $J_2=1$), which is close to the critical point and the nature of which is controversial according to previous studies.
The strip geometry has open boundary conditions along both the $x$- and $y$-directions. 
The DMRG simulations with spin SU(2) symmetry use a maximum bond dimension of $M=10000$ SU(2) multiplets, which is equivalent to about $40000$ U(1) states and ensures a good convergence with the truncation error at about $1\times 10^{-6}$. 
The ground-state energy per site for each $M$ is listed in the legend of Fig.~\ref{fig:DMRG}(a).
The DMRG correlation functions are measured along the central row ($y = 6$) and are extrapolated versus $1/M$ ($M=4000-10000$) following a second order polynomial fit.
The extrapolated results are shown by the green filled circles in Figs.~\ref{fig:DMRG}(a) and \ref{fig:DMRG}(c). 
For the spin correlation function $\langle {\bf S}_0 \cdot {\bf S}_r \rangle$, the reference site ${\bf S}_0$ is chosen at the third column near the left edge.
For the dimer correlation function $\langle B_0 B_r \rangle - \langle B_0 \rangle \langle B_r \rangle$, the bond operators are chosen along the $x$ direction, namely $B_r = {\bf S}_r \cdot {\bf S}_{r + \hat{x}}$ ($\hat{x}$ is the unit vector along the $x$ direction).
The reference bond for dimer correlations is chosen between the third and fourth columns near the left boundary.

The PEPS results from $D=4-10$ are also presented to compare with the DMRG results. Firstly, we note that the PEPS energies at $D=8$ and $10$ agree very well with the $M=10000$ DMRG energy. 
Meanwhile, the converged PEPS correlations also agree very well with those obtained from DMRG, as shown in Figs.~\ref{fig:DMRG}(b) and \ref{fig:DMRG}(d). 
These comparisons further demonstrate the accuracy of our PEPS results. 
It also shows that the PEPS simulations with $D=8$ are good enough to  converge the results for the $12 \times 28$ SSM. 

Here we explicitly show the convergence of energy and order parameters on the largest available  $20\times 20$ SSM system at $J_1/J_2=0.8$, shown  in Table.~\ref{tab:convergence}. In the 
 calculations, for the order parameters $\langle M^2 \rangle$ and $\langle Q^2 \rangle$ defined based on correlation functions, all pairs of spin and dimer correlations are computed for summation, respectively. Other order parameters  mainly involve local terms, which have less computational cost and can be obtained with smaller sampling uncertainty, allowing to check the convergence in higher accuracy.  Similar to $12\times 28$, for $20\times 20$ we can see visible improvement by increasing $D=4$ to $D=8$, and the $D=6$ results  are rather close to $D=8$ results.   Further increasing $D$ to 10, the results almost have no further visible improvement, indicating $D=8$ indeed converges well for the physical quantities.  Actually, it has also been shown that the $D=8$ PEPS simulations can converge quite well for other highly frustrated spin models~\cite{liuQSL,liuj1j2j3,j1xj1yj2}, including the $J_1$-$J_2$-$J_3$ model up to $20\times 28$ sites~\cite{liuj1j2j3}, as well as the above highly nontrivial $J-Q$ model. In the meanwhile, in Table.~\ref{tab:convergence}, we can see for large $D$ it has $\langle M^2 \rangle=3\langle M^2_z\rangle$ and $\langle M_z \rangle $ is almost zero within the Monte Carlo sampling error where $M_z=\frac{1}{L^2}\sum_{i_x,i_y}(-1)^{i_x+i_y}S^z_{i_x,i_y}$, indicating the SU(2) symmetry of the ground state  recovers very well.

\begin{table*}[htb]
   \centering
 \caption { The convergence of energy and order parameters  with respect to the PEPS bond dimension $D$ on the $20\times 20$ SSM system at $J_1/J_2=0.8$ (set $J_2=1$). $\langle P \rangle^2 $, $\langle D \rangle^2 $ and $\langle Q^2 \rangle $ are all valid VBS order parameters, used for cross check;  $\langle M^2 \rangle $ are AFM order parameters, and  3$\langle M^2_z \rangle$ and $\langle M_z \rangle$ are used to check the SU(2) symmetry.
 }
	\begin{tabular*}{\hsize}{@{}@{\extracolsep{\fill}}cccccccc@{}}
		\hline\hline
	   $D$  & $E$ & $\langle P \rangle^2 $ & $\langle D \rangle^2 $   & $\langle Q^2 \rangle $  & $\langle M^2 \rangle $& 3$\langle M^2_z \rangle$  & $\langle M_z \rangle$  \\ \hline
 		
        4 & -0.442104(8) & 0.01624(8) & 0.01738(9) & 0.0380(2) &0.0139(1)&0.01450(9)& 0.0014(4) \\
        6 & -0.442861(9) & 0.01349(5) & 0.01454(5)  & 0.0324(1) &0.0158(1)&0.01571(9)&0.0011(5)\\
        8 & -0.442920(2) & 0.01341(8) & 0.01442(9)  & 0.0321(2) &0.0155(1)&0.01548(6)&0.0001(5)\\  
        10 &-0.442928(3) &0.01344(9)  &0.01441(5)  &0.0320(1) & 0.0155(2) &0.01549(5)&0.0003(4)\\ 
 		\hline\hline
	\end{tabular*}
\label{tab:convergence}	
\end{table*}

 \subsection{comparison with iPEPS and iPESS}

\begin{figure}[htbp]
 \centering
 \includegraphics[width=3.4in]{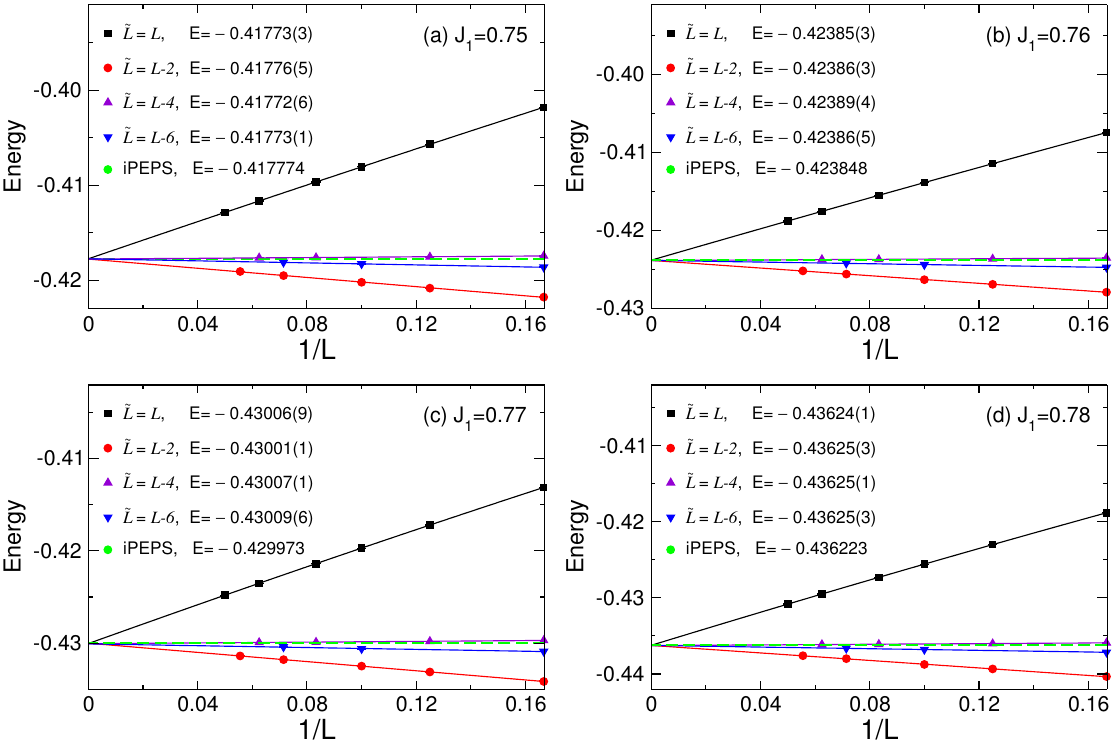}
 \caption{ Finite size scaling of ground state energies with $L=6-20$ for SSM at different $J_1$. The iPEPS results are from Ref.~\cite{corboz2013tensor}.  For $\tilde{L}=L $ second-order polynomial fits are used and for other cases, linear fits are used. Extrapolated thermodynamic limit energies are listed in the legend. }
 \label{fig:energy}
 \end{figure}

We conduct a comparison of energy per site between the finite PEPS and iPEPS methods, shown in Fig.~\ref{fig:energy}. For this purpose, we utilize the best iPEPS energies reported in Ref.~\cite{corboz2013tensor} at $J_1/J_2=0.75$, 0.76, 0.77, and 0.78, all obtained with a bond dimension of $D=10$.
To obtain the thermodynamic limit energy for our finite size calculations, we perform an extrapolation of the finite PEPS energy using systems with sizes ranging from $L=6$ to $L=20$. To facilitate finite-size scaling, one can choose different central cluster $\tilde{L} \times \tilde{L}$ for finite-size scaling~\cite{liufinitePEPS,liuQSL,liuj1j2j3},  as depicted in Fig.\ref{fig:energy}. Regardless of the chosen central cluster, the extrapolated energies for each $J_1/J_2$ remain almost identical, similar to previous studies~\cite{liufinitePEPS,liuQSL,liuj1j2j3}. 
These extrapolated PEPS energies and iPEPS energies show a nice agreement with up to 5 significant digits. 

\begin{figure}[htbp]
 \centering
 \includegraphics[width=3.4in]{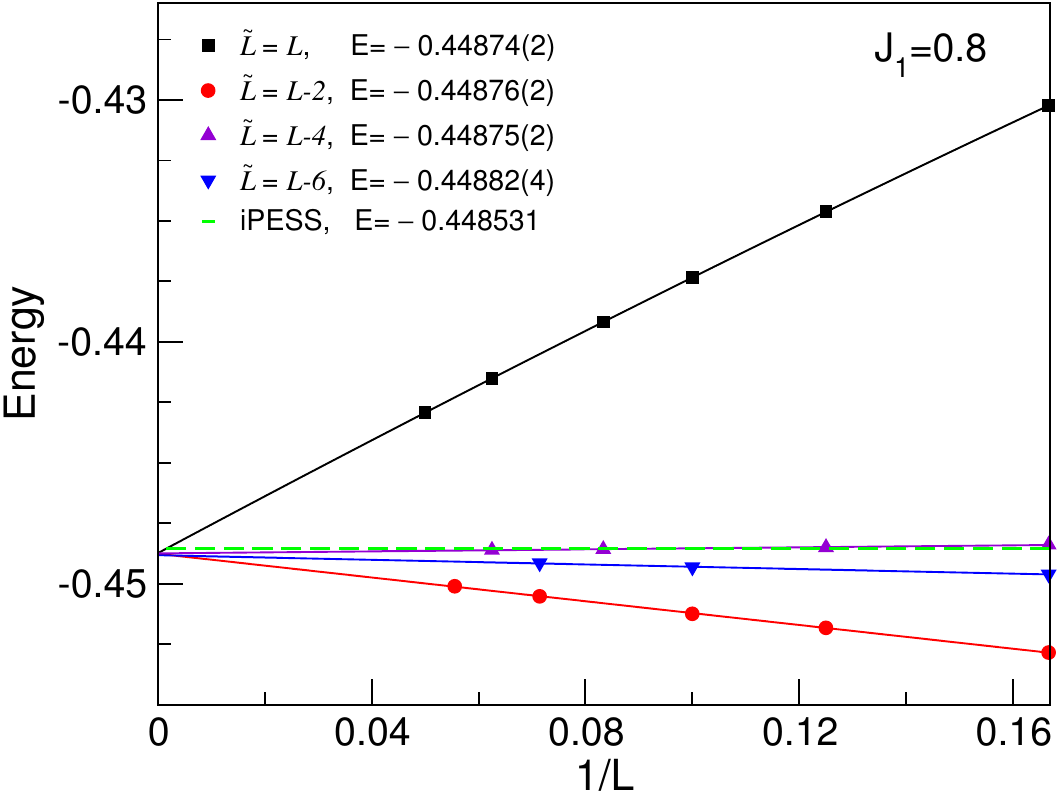}
 \caption{ Finite size scaling of ground state energies with $L=6-20$ for SSM at $J_1=0.8$. The iPESS results are from Ref.~\cite{xi2021first}.  For $\tilde{L}=L $ second-order polynomial fits are used and for other cases, linear fits are used. Extrapolated thermodynamic limit energies are listed in the legend. }
 \label{fig:energy_J1_0.8}
 \end{figure}

We also compare the SSM energies at $J_1/J_2=0.8$ with other available results, here from the infinite-size projected entangled simplex states (iPESS) results, shown in Fig.~\ref{fig:energy_J1_0.8}. The best iPESS result ($D=7$) is taken from  Ref.~\cite{corboz2013tensor}. We can see the extrapolated finite PEPS energies also agree well with the iPESS results up to 4 significant digits.

  \begin{figure}[htbp]
 \centering
 \includegraphics[width=3.4in]{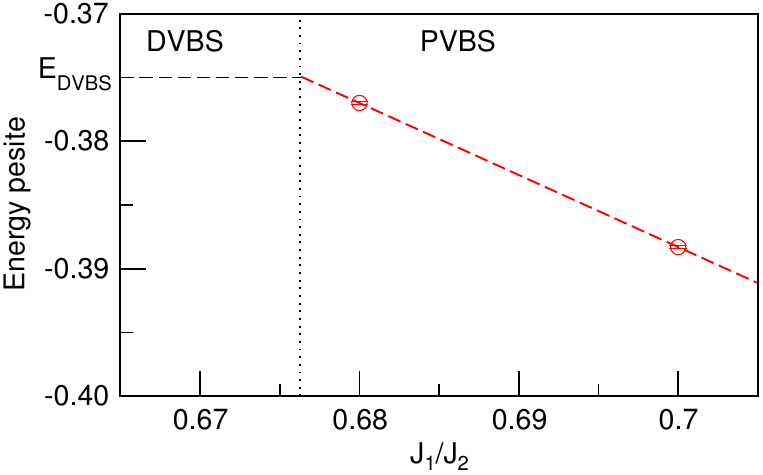}
 \caption{Energy variation with respect to $J_1/J_2$ for the SSM. Red points denote thermodynamic limit energies at $J_1/J_2=0.68$ and 0.72, and red dashed line denotes a presumed linear change of the energies close to the first-order transition point. }
 \label{fig:energy_DVBS}
 \end{figure}

Similar to the iPEPS analysis in Ref.~\cite{corboz2013tensor}, we can determine the first-order transition point between the DVBS (dimer valence-bond solid) and PVBS (plaquette valence-bond solid) phases by examining the ground state energy. The DVBS phase corresponds to a dimer state with an exact energy $E_{DVBS}=-0.375$.
Using finite size scaling,  we can get the thermodynamic limit energies at $J_1/J_2=0.68$ and 0.72, and estimate the energies at other $J_1/J_2$ points with a presumed linear extrapolation with respect to $J_1/J_2$ (a linear assumption is reasonable as the energy seems to change linearly  in a large range of $J_1/J_2$ according to our results or previous studies~\cite{corboz2013tensor,lee2019signatures,xi2021first} ).  
 The energy variation about $J_1/J_2$, obviously suggests a first-order DVBS-PVBS transition at  $J_1/J_2\simeq 0.676$, shown in Fig.~\ref{fig:energy_DVBS}, which is also in good agreement with the iPEPS and iPESS results, as well as the DMRG  study~\cite{lee2019signatures}.

\begin{figure}[tbp]
 \centering
 \includegraphics[width=3.4in]{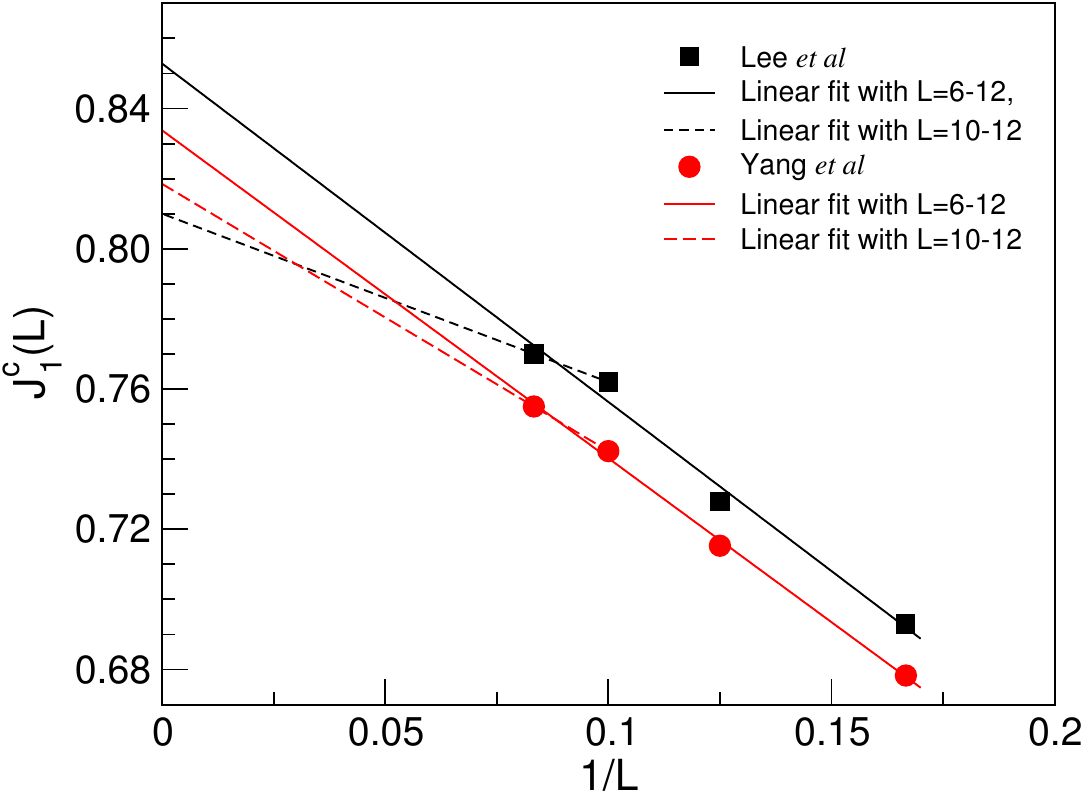}
 \caption{Finite-size extrapolations of the relevant points of the DMRG results from Ref.~\cite{lee2019signatures} (black) and Ref.~\cite{yang2022quantum} (red). They use different systems and physical quantities to locate the turning points, so their values for a given cylinder width $L$ could be different.  Linear fits with $L=6-12$ and $L=10-12$ are shown, which give extrapolated values 0.85(1) and 0.81 using the results from Ref.~\cite{lee2019signatures} and, 0.834(4) and 0.818 using singlet minimum results from Ref.~\cite{yang2022quantum}, respectively.  }
 \label{fig:iDMRG}
 \end{figure}

\subsection{PVBS-AFM transition point}
 We now examine the PVBS-AFM phase transition point in the SSM determined by various methods, which is very challenging for accurate determination. The outcomes of the DMRG investigation conducted on an infinitely long cylinder ($L_x$ is infinite while $L_y=L$ is finite) show apparent finite-size effects~\cite{lee2019signatures}. The observed values of $J^c_1(L)$ are 0.693, 0.728, 0.762, and 0.77 for $L$ equaling 6, 8, 10, and 12, respectively~\cite{lee2019signatures}. Extrapolating to the 2D limit, the transition point is estimated as $J^c_1=0.85(1)$ (linear fit $1/L$ with $L$ ranging from 6 to 12) or $J^c_1=0.81$ (linear fit $1/L$ with $L$ ranging from 10 to 12), as illustrated in Fig.~\ref{fig:iDMRG}. On the other hand, based on the singlet minimum points from the later DMRG study~\cite{yang2022quantum} ($L_x=2L$ is finite), the estimation of $g_{c1}$ with similar linear fits can also be 0.834(4) for $L=6-12$ and 0.818 for $L=10-12$ [see Fig.~\ref{fig:iDMRG}], rather close to $g_{c2}\approx 0.82\sim 0.83$~\cite{yang2022quantum,wang2022quantum}. This indicates the possibility that $g_{c1}$ and $g_{c2}$ coincide, and the proposed narrow spin liquid might be a consequence of finite-size effects.
Noticeably, these estimations based on different cylindrical systems and different physical quantities, seemingly reach an point around $J_1\sim 0.82$ in the 2D limit, closely resembling our obtained PVBS-AFM transition point, $J^c_1=0.828(5)$.

Additionally, the PVBS-AFM phase transition points from infinite tensor network methods are a bit different from our $J^c_1=0.828(5)$, which are  $J^c_{1}=0.765(15)$ \cite{corboz2013tensor} or 0.79 \cite{xi2021first}.  Generally speaking,  finite PEPS methods suffer finite size effects (the finite-$D$ effects can be removed by a mild $D$  for a moderate system  size, e.g. here $D=8$ for $20\times 20$), and infinite tensor network methods suffer finite $D$ effects. 
 In our finite-size calculations, the uncertainty caused by finite-size effects for the PVBS-AFM transition point has been significantly reduced by crosschecking different physical quantities including the order parameter ratio that is related to emergent symmetry.  For infinite-size tensor network simulations,  we believe a proper consideration about finite $D$ effects may also give the similar result. We point out that for larger finite-size systems larger $D$ is necessary for convergence, and the systematical investigation of finite-$L$ and finite-$D$ effects can be found in Refs.~\cite{huang2023tensor,ueda2023finite}.

\section{Setup for open boundary systems}

In open boundary systems, the presence of boundaries will induce dimerized patterns, which requires careful consideration about such effects on the real plaquette pattern in the PVBS phase. To illustrate this, we examine the Shastry-Sutherland model (SSM) and calculate the values of local plaquette operators at each site.

\begin{figure*}[htbp]
 \centering
 \includegraphics[width=6.8in]{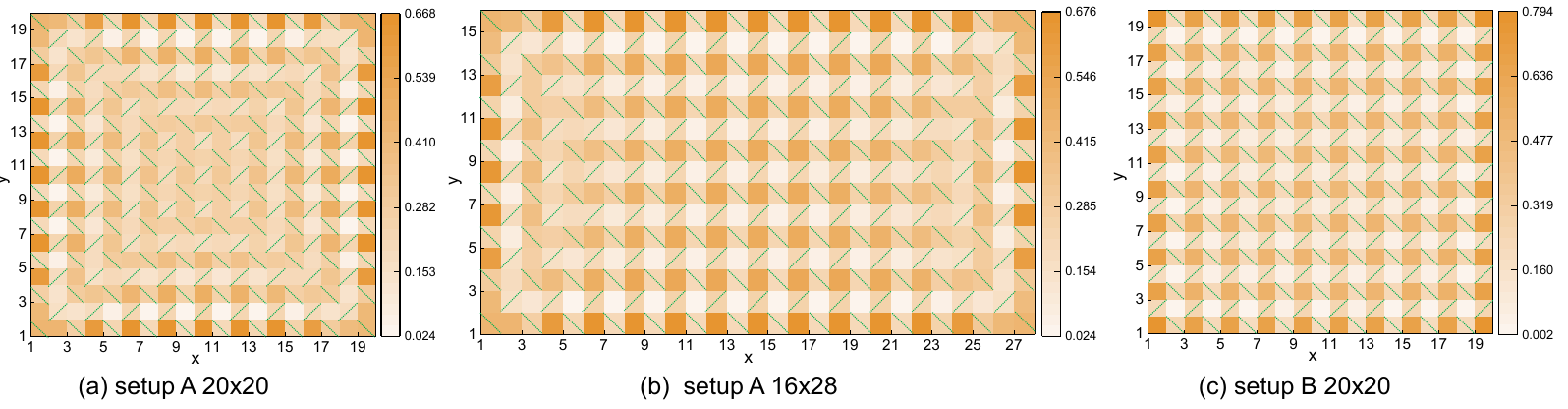}
 \caption{ Two setups for the SSM. Green diagonal line denote diagonal interaction terms. Corner sites have diagonal interaction terms in setup A and  but not in setup B. (a-b) show the plaquette operator values $\Pi_{\bf i}$ setup A on $20\times 20$ and $16\times 28$ systems. (c) shows the plaquette values on $20\times 20$ using setup B. All three cases use $J_1/J_2=0.74$.  }
 \label{fig:setup}
 \end{figure*}

We start by considering setup A, where diagonal interaction terms exist at the corners. We consider an $L\times L$ lattice with $L=20$. We observe a distinct plaquette  pattern near the boundaries of the systems, but it becomes blurry in the middle region, shown in Fig.~\ref{fig:setup}(a). This blurriness is a consequence of the boundaries. Specifically, the $x$-directional boundaries induce a certain pattern, while the $y$-directional boundaries give rise to another kind of plaquette pattern. The 'superposition' of these equal-footing boundary effects results in the formation of domain walls along the diagonal directions [see Fig.~\ref{fig:setup}(a)].
However, if the $x$- and $y$-boundaries are not treated on equal footing, a clear plaquette pattern may eventually emerge in the middle region, as illustrated in Fig.~\ref{fig:setup}(b) for a $16\times 28$ system. 

In setup A, the presence of domain walls poses a challenge for finite-size scaling since the VBS order parameters defined on the entire lattice are not well-defined due to the incompatibility of the plaquette pattern in different domains. 
For example, considering the boundary-induced plaquette order parameter $P=\frac{1}{N_p}\sum_{{\bf i}}(-1)^{i_x}\Pi_{{\bf i}}$, the contributions from different domains could be partly canceled (so is the VBS order parameters based on bond-bond correlations).  Even if one attempts to define the VBS order parameters within a specific domain, the ambiguity caused by the domain walls cannot be entirely removed. 

The plaquette pattern in the middle region of  $16\times 28$ system using setup A [Fig.~\ref{fig:setup}(b)], suggests one should use the setup B, where corner sites have no diagonal interaction terms. In this context, the plaquette pattern is compatible with the boundaries [see Fig.~\ref{fig:setup}(c)]. As a result, there are no domain walls, and the VBS order parameters can be defined unambiguously over the lattice, which ensures a more reliable and consistent analysis of the VBS order parameters. Consequently, throughout this work, setup B is used to compute all physical quantities in the SSM and the extended SSM.

\bibliography{mainl}

\end{document}